\documentclass[runningheads]{llncs}

\usepackage{todonotes}
\usepackage{fullpage}
\usepackage{amsfonts,amssymb,amsmath}

\usepackage{xspace}
\newcommand{\etal}{\textit{et al.}\xspace}
\newcommand{\etc}{\textit{etc.}\xspace}
\newcommand{\ie}{\textit{i.e.,}\xspace}
\newcommand{\eg}{\textit{e.g.,}\xspace}

\newcommand{\draft}[1]{#1}

\newcommand{\C}{{\mathcal C}}
\newcommand{\Z}{{\mathbb Z}}
\newcommand{\F}{{\mathbb F}}
\newcommand{\R}{{\mathbb R}}
\newcommand{\N}{{\mathbb N}}

\renewcommand{\AA}{\mathcal{A}}

\newcommand{\OO}{\textsf{Obf}}
\newcommand{\Obf}{\textsf{Obf}}
\newcommand{\PP}{{\mathcal P}}

\newcommand{\asset}{\textsf{a}}
\newcommand{\aux}{\textsf{aux}}
\newcommand{\Gen}{\textsf{Gen}}
\newcommand{\verify}{\mathcal{V}}
\newcommand{\Sample}{\textsf{Gen}}

\begin{document}

\title{Towards a Theory of Special-purpose Program Obfuscation}



\author{Muhammad Rizwan Asghar\inst{1} \and
Steven D. Galbraith\inst{2} \and
Andrea Lanzi\inst{3} \and \\
Giovanni Russello\inst{1} \and
Lukas Zobernig\inst{1}
}
\authorrunning{Asghar \etal}
%
\institute{School of Computer Science, The University of Auckland, New Zealand \and
Mathematics Department, The University of Auckland, New Zealand
\and
Computer Science Department, Universita degli studi di Milano, Italy\\
}

\maketitle

\begin{abstract}
Most recent theoretical literature on program obfuscation is based on notions like Virtual Black Box (VBB) obfuscation and indistinguishability Obfuscation (iO).
These notions are very strong and are hard to satisfy.
Further, they offer far more protection than is typically required in practical applications.
On the other hand, the security notions introduced by software security researchers are suitable for practical designs but are not formal or precise enough to enable researchers to provide a quantitative security assurance.
Hence, in this paper, we introduce a new formalism for practical program obfuscation that still allows rigorous security proofs.
We believe our formalism will make it easier to analyse the security of obfuscation schemes.
To show the flexibility and power of our formalism, we give a number of examples.
Moreover, we explain the close relationship between our formalism and the task of providing obfuscation challenges.

This is the full version of the paper. In this version we also give a new rigorous analysis of several obfuscation techniques and we provide directions for future research.
%
\end{abstract}

\section{Introduction}

The goal of program obfuscation is often informally stated as ``to hide semantic properties of program code from a Man-At-The-End (MATE) attacker who has access to an executable file of the program''.
\footnote{The phrase ``executable file'' does not only mean binary code. It can mean code in any language that is naturally provided for execution on a device, such as Java bytecode.}
This is an imprecise notion, and despite more than twenty years of research in software security that has led to a variety of obfuscation tools and techniques, there is still not a useful and practical formalism for what it means to ``hide semantic properties of program code''.

In 1996, Goldreich and Ostrovsky~\cite{GO96} suggested that obfuscation should prevent ``the release of any information about the original program which is not implied by its input/output relation and time/space complexity''.
In 1997, Collberg, Thomborson, and Low~\cite{CTL97} remarked that ``developers are mostly frightened by the prospect of a competitor being able to extract proprietary algorithms and data structures from their applications'' and stated ``Informally, a transformation is potent if it does a good job confusing Bob, by hiding the intent of Alice's original code.''

The landmark paper by Barak \etal \cite{BGIRSVY01,BGIRSVY12} in 2001 introduced the notion of Virtual Black Box (VBB) obfuscation.
Informally, a program is VBB obfuscated if an attacker can learn nothing more, when given the executable code, than what could be learned from running it (\ie oracle access). 
This is a very strong notion that captures the intention of ``hiding semantic properties of program code''.
But unfortunately, Barak \etal show that this notion is impossible to achieve in general, by showing that for every potential obfuscating compiler $O$, there is some program $P$ and an attacker $A$ such that given $\OO(P)$, the attacker can learn something about $P$ that cannot be learned just from the input-output behaviour of $P$.

The work of Barak \etal (and also much of the work in the software security community)
aims to produce \emph{general-purpose} obfuscation techniques.
This is a worthwhile goal, but in our opinion, it is too ambitious.
In contrast, there is also a very successful theme in obfuscation research to consider specific classes of programs and to give obfuscation methods that are designed with a specific goal in mind.
We call this \emph{special-purpose obfuscation}.
The classic example of this is programs that perform a licence check or password check, also called \emph{point functions}. 
A point function is a program that takes an input $x$ from the user and checks whether $x$ is equal to some particular hard-wired value $c$ (say the licence key or password) and only continues execution if the equality holds. 
The goal of the obfuscating compiler is to hide the value $c$, so that it cannot be extracted from the obfuscated executable code by any kind of reverse engineering tool.
A standard solution to this problem is to use a cryptographic hash function $H$ and to store $c' = H(c)$ in the obfuscated executable. 
The obfuscated program then computes $H(x)$ and checks if it is equal to $c'$.
This research avenue has been greatly extended under the theme of \emph{evasive functions}, which we discuss in detail in Section~\ref{sec:evasive}.

As discussed by Collberg~\cite{Col18}, program obfuscation still does have a number of practical applications.
Hence, there is a need to find practical solutions.

Several authors have recognised the need for more precise and formal approaches to practical obfuscation.
Kuzurin \etal~\cite{KSVZ07}, in their discussion of solutions from the software security community, state ``The principal drawback of all known obfuscation techniques is that they do not refer to any formal definition of obfuscation security and do not have a firm ground for estimating to what extent such methods serve the purpose.''
They go on to say
``We believe that further progress in the study of program obfuscation may ensue primarily from the development of a solid framework which makes it possible to set up security definitions for program obfuscation in the context of
various applications''.
Similarly, Xu \etal~\cite{XZKL17} compare applied (they call it ``code-oriented'') and theoretical (named ``model-oriented'') obfuscation approaches.
They claim ``current evaluation metrics are not adequate for security purposes''.
Their conclusion urges ``the communities to rethink the notion of security and usable program obfuscation and facilitate the development of such obfuscation approaches in the future.''

The goal of this paper is to respond to these calls and to propose a useful formalism for obfuscation.

\paragraph{Our results}
The aim of this paper is to introduce a formalism for program obfuscation that achieves three main goals:
\begin{enumerate}

\item  It is powerful enough to capture real-world applications of program obfuscation.

\item  It is precise enough to allow formal analysis of obfuscation techniques and to prove their security.

\item  It bypasses impossibility results and is plausible to be achieved.

\end{enumerate}

Our formalism uses the notion of \emph{assets} of a class of programs.

The history of research on both theoretical and applied obfuscation indicates that general-purpose obfuscation is too hard to achieve.
Further, general-purpose obfuscation is probably unnecessary for most practical applications. 
In practice, software developers usually have a particular secret or intellectual property that they wish to protect, rather than desiring to secure ``all semantic properties'' of a program.
Hence, we believe it is more effective to develop special-purpose obfuscators that are targeted to protect a \emph{particular type} of asset in a \emph{particular class} of programs.

The rest of this paper is organised as follows.
In Section \ref{sec:related}, we review related work.
Our formalism is presented in Section~\ref{sec:formalism}. 
Two of our contributions are to define what it means to specify a class of programs, and to define the correctness of an attacker that learns an asset.
One theme that we want to stress is the connection between formalising security and providing obfuscation challenges.
We discuss this connection in Section~\ref{sec:obfuscation-challenges}.
Section \ref{sec:act} illustrates how our formalism is intended to be used. 
The goal of Section~\ref{sec:examples} is to discuss how various real-world examples can be modelled within our formalism.
Please note that Section~\ref{sec:examples} is not intended to be a complete list of all possible programs and that this section is not giving new obfuscation techniques or protection mechanisms.
We stress that we do not claim that our formalism includes all possible program types or security notions.
In particular,
\begin{enumerate}
\item We do not claim that our formalism includes all possible program types or all possible security notions or security goals.
\item We do not claim our list of examples is complete.
\item  We do not give new obfuscation solutions for all the examples in the paper.
\end{enumerate}
We hope our formalism will be a useful starting point to address some of these open problems, and that it will be further extended by the research community.

In Section \ref{sec:composition}, we discuss the security of applying different obfuscators iteratively on a program.
We discuss limitations of our formalism in Section~\ref{sec:limitations}.
Finally, we conclude this paper in Section \ref{sec:conclusion}.

\section{Previous Work on Obfuscation Definitions}
\label{sec:related}

Goldreich and Ostrovsky~\cite{GO96} suggested a notion of obfuscation security with a similar flavour to the later notion of \emph{VBB obfuscation} defined by Barak \etal \cite{BGIRSVY01}.
The theoretical community has continued to mainly use such definitions.
In the software security field, there has been a broader range of notions discussed, some of them imprecisely. 
We briefly survey them below.

Collberg, Thomborson, and Low~\cite{CTL97}  discuss general software engineering measures of code complexity.
\draft{Chapter 4 of Collberg and Nagra~\cite{CN09} gives a more formal definition of potency (Definition 4.3) but does not define resilience; they state ``In our original paper on obfuscation, we proposed that an obfuscating transformation should be evaluated in terms of its potency, resilience, cost, and stealth'' but then they deprecate some aspects of these definitions ``you still occasionally see references to these definitions in the literature''.
}
Several other papers have discussed complexity measures from software engineering (\eg Anckaert \etal in \cite{AMSBBP07}).

\draft{White-box cryptography~\cite{CEJO02} is a special case of obfuscation, where the program being obfuscated is usually a symmetric encryption or decryption algorithm and one purpose of the obfuscation is to hide the secret key.
Saxena, Wyseur, and Preneel~\cite{SWP09} discuss security notions for white-box cryptography.
In their work they describe a strong notion that is more-or-less the same as VBB security and give some impossibility results.}

The notion of asset has been discussed in several papers, as we now explain.

Basile \etal \cite{BDCHNW09} introduce the notion of ``business goals'' of a program.
The ``defender associates many business goals to software'' and a successful attack means ``business goals not satisfied anymore''.

In 2014, Barak \etal \cite{BBCKPS14} define \emph{input-hiding obfuscation} for point functions and evasive functions: Given an obfuscated program, it should be hard for an attacker to find an input $x$ that the program accepts.
They showed that this notion is not always implied by VBB security when the set of accepted inputs is super-polynomially large.

In Chapter 5 of~\cite{CN09}, Collberg and Nagra give the notion of an \emph{asset}, specified by programmer/developer.
Their work also restricts focus in each example to a class ``Input Programs''.
Unfortunately, Definition~5.2 of~\cite{CN09} is unclear: ``An asset is a derivable property of a program $P$ and its set of inputs $I$ such that asset$(P,I) = 1$''.
That is, the term ``derivable property'' is not defined, and the question of what it means to specify the class ``Input Programs'' is not made precise.
The chapter gives some examples of assets, such as (page 306) ``$P$ never outputs $x$''.
Nevertheless, the book overall is written in a friendly and conversational manner, rather than in the formalism of theoretical cryptography.

Ahmadvand, Pretschner, and Kelbert (see Section 3.1 of~\cite{AhPrKe2018}) define assets to be ``elements whose integrity needs to be protected against attacks'' and ``valuable data or sensitive behaviour(s) of the system, tampering with which renders the system's security defeated''.

Schrittweiser \etal \cite{SKKMW16}, in Section 2.3 of their impressive survey paper, 
give a selection of assets phrased in terms of attack goals. 
Their work illustrates these goals in the context of Digital Rights Management (DRM) applications. 
The list of goals includes: finding location of data (such as a secret key); finding location of program functionality (such as the entry point of a copy-protection mechanism); extraction of code fragments (the media decryption algorithm); and understanding the program.

To summarise, many authors have proposed a security model that involves the notion of an asset. But none of the previous works gives a precise enough formalism to enable a rigorous cryptographical approach to the problem. 
The main goal of our paper is to present such a formalism.


Note that the choice of asset depends on the context, and so a particular asset only makes sense within some class of programs.
Hence, it is appropriate to consider \emph{special-purpose obfuscation tools}, tailored for certain types of asset.
This idea is implicit in much of the theoretical research on obfuscation.
An explicit argument in favour of special-purpose obfuscation is given in
Section 3.4 of~\cite{HMLS07}: ``What we advocate here is to consider specific obfuscators for specific function families''.

\section{Our Formalism} \label{sec:formalism}

Following Collberg and Nagra~\cite{CN09}, we take a developer-centered approach to defining security of an obfuscator.
Building on~\cite{AhPrKe2018,BDCHNW09,BBCKPS14,CN09}, we work with the notion of an asset and define security in a similar way to the definition of potency given as Definition 4.3 in~\cite{CN09}.
We do not concern ourselves in this paper with stealth (\ie can an attacker determine that some form of obfuscation has been applied to the program segment) because, in most non-malware contexts, it will be public knowledge that obfuscation is being used.

Also, following much of the theoretical work, we restrict attention to \emph{classes of programs} rather than trying to formulate definitions that work for any circuit or any Turing machine.
Two of our contributions are to define what it means to specify a class of programs, and to define the correctness of an attacker that learns an asset.

\draft{Following the tradition in theoretical cryptography, we do not formalise security against a human attacker. 
In practice, it is hard to prove that an obfuscation technique cannot be broken by a human.
Instead, w}
We assume that the adversary is a Turing machine, so that we can use the standard theoretical tools of security reductions.

We will evaluate the threat of an attacker in terms of its running time. 
To be able to make complexity-theoretic statements (\eg the attacker is polynomial time or exponential time), there must be a parameter that takes arbitrarily large values with respect to which the running time is expressed.
It does not make sense to think of an ``exponential-time attack'' against a single program.
Hence, we need to parameterise program classes with a security parameter $n$.

There are three entities in our formalism:
\begin{itemize}

\item \textbf{Programmer/developer:} 
A human who has developed a program (written in some high-level source code) that contains some asset or intellectual property. 

\item \textbf{Obfuscator/defender:} 
A randomised algorithm that takes as input a program in some format and outputs a program in some format. 
Note that the input format may not be same as output format, \eg a compiler is an obfuscator according to some research papers. 
We study \emph{special-purpose obfuscators} that depend on the particular program class and asset.
\draft{Indeed, a large amount of the previous work on obfuscation can be viewed as protecting a certain asset in a certain class of programs.}

\item \textbf{Attacker/de-obfuscator:} 
An algorithm that takes program code in some format and tries to learn the asset in the program.
\end{itemize}

We now state our formalism, step by step.

\subsection{Program (Segment) Classes}

\begin{definition}
A \textbf{program class} (or \textbf{class of program segments}) $\C$ is a set of programs, all written in the same language.
For each $n \in \N$, there is a (possibly empty) subset $\C_n \subseteq \C$.
The sets $\C_n$ are disjoint and $\C = \cup_{n \in \N} \C_n$.
There is a polynomial $p(x)$ such that for all $n \in \N$ and all $P \in \C_n$ the running time of $P$ on any input is bounded by $p( n )$.
Further, we require that the class $\C$ has a compact description in the following sense:
\begin{enumerate}
\item There is an efficient program generator $\Gen_\C$ for the class $\C$.
Formally, this is a randomised polynomial-time algorithm that on input $1^n$ (a bitstring of length $n$) computes in time bounded by a fixed polynomial in $n$ a random program $P \in \C_n$ and some auxiliary data $\aux$.
The auxiliary data is typically related to assets in the program, and may include information that is needed by the obfuscator and/or to verify the asset.

\item $\C_n$ should asymptotically be super-polynomial in size, so for all polynomials $p(n)$ we have $ \# \C_n \ge p(n)$ for sufficiently large $n$.

\end{enumerate}
\end{definition}

Here, $n$ acts as a security parameter, and the sets $\C_n$ are intended to consist of programs with a similar security level in terms of finding an asset. 

The requirements on $\C_n$ are implicit in the previous literature on theoretical obfuscation, but we believe it is worthwhile to make them explicit.
For example, Barak \etal \cite{BGIRSVY01} define obfuscation for ``a family of Turing machines'' and ``a family of circuits'', without explaining how such a family is represented. 
They define a security experiment that implicitly assumes the ability to sample Turing machines/circuits from the family. We have made this explicit with the algorithm $\Gen_C$.

We could consider $\C$ to be a set of circuits or Turing machines.
But in practice, it will be more convenient to think of a set of instructions in some programming language.
The formalism can be applied to programs in any code.
For example, the class $\C$ could comprise programs in source code written by the developer, or bytecode, or compiled native code.

In practice, program classes are defined by software developers (we will give examples in Section~\ref{sec:examples}) so that they have some meaningful common feature or functionality. 
The parameter $n$ has some natural meaning to the developer, such as bit-length of user input or key-length of some secret that is stored in the program.
We stress that we focus on program segments, since a large software project will have many components and the type of intellectual property and protection level required may vary among them.

The requirement that $\C$ is defined by a random program generator may seem counter-intuitive.
%
However, we view this as a kind of Kerckhoff's principle. 
The crucial point is that the hard part of software development is finding the exactly right program $P \in \C_n$ for the problem at hand. 
So, a software developer can easily generate a random program $P \in \C_n$, but the probability to have the asset must be low.

\subsection{Assets}

The formulation of asset is non-trivial, since the set of all possible assets might be exponentially large (\eg the set of all inputs that are accepted by some evasive function), or there may be many different representations for an asset (\eg different representations of the same control flow graph).
Our insight is that the focus should be less on what is an asset and more on how the software developer recognises that the asset has been successfully found by the adversary.

\begin{definition}
Let $\C = ( \C_n )$ be a class of programs.
An \textbf{asset space} $\AA$ for $\C$ is a family of sets $(\AA_n)_{n \in \N}$.
An \textbf{asset} for $\C$ is a sequence of functions $\asset_n : \C_n \to \PP( \AA_n )$, where $\PP(\AA_n)$ is the power set of $\AA$.
(This function does not need to be efficiently computable.)

An asset is \textbf{efficiently verifiable} if there is an algorithm $\verify$ (the \textbf{asset verifier}) and a polynomial $q(x)$ such that for all $n \in \N$ and all $(P, \aux) \leftarrow \Sample_\C(n)$ (so $P \in \C_n$), and all $a \in \AA_n$, $\verify( P, \aux, a ) $ outputs $1$ if $a \in \asset_n(P)$ and $0$ otherwise, the running time of $\verify$ on $\C_n$ is bounded by $q( n )$.
\end{definition}

The above definition has a deterministic verifier that always outputs the correct answer.
But one can also consider a randomised verifier that is correct with some overwhelming probability; this can arise in some settings.
We have given $\verify$ the original code of $P$ and the auxiliary information in order to check the asset, but in many applications, oracle access to $P$ would suffice.

Finally, we can discuss the attacker and the security goal.
We follow Kerckhoffs's principle and so we assume the attacker knows: 
the class $\C$ of programs being obfuscated; the asset space $\{ (\AA_n, \asset_n) \}$; and the obfuscation tool being used to protect the asset.
Hence, our formalism does not address the question of \emph{stealth}, \ie can an attacker determine that part of a program has been obfuscated, and if so, using what tool?
Indeed, we believe that highly secure obfuscation and stealth are probably incompatible.

Assets should not be guessable or easily \emph{learnable} from the program.\footnote{This is a point of difference between our formalism and notions like VBB and indistinguishability Obfuscation (iO): In those settings, a learnable asset is considered trivially obfuscatable (since the obfuscated program hides everything that is not learnable from input/output behaviour); whereas, in our formalism, such assets are un-obfuscatable.}
If the asset is learnable from executing the program on various inputs (and logging those inputs and corresponding outputs) then it is clearly impossible to protect the asset using any form of obfuscation.
Hence, we require $\# \AA_n$ to be large.

For example, the class of programs that computes linear functions $\C_n = \{ L : \F_q^n \to \F_q^m \}$ is learnable by executing such a program on unit vectors in $\F_q^n$. 
Hence, it makes no sense to obfuscate $\C_n$.
There are many program classes that might be tempting to obfuscate, but are actually efficiently learnable. 
For example, consider a predicate $P$ that tells whether or not an input $x \in \Z$ satisfies $x < C$ for some secret constant $C$. 
One can learn $C$ using black-box access to $P$ by binary search.

It is the software developer's job to define ``useful'' or ``meaningful'' assets.
There may be some particular programs $P \in \C_n$ that are easily learnable, but we require that a random program in the class is not learnable.
More precisely, the output distribution of $\Gen_\C$ needs to be such that, for large $n$, instances output by $\Gen_\C(n)$ are not learnable with high probability.


\subsection{Secure Obfuscators}

We now define what is meant by an obfuscator. 
First, it is a compiler. 
So, it takes programs in some language and converts to another form.

\begin{definition} \label{defn:secure-obfuscator}
Let $\C = (\C_n)$ be a class of programs defined by a generator $\Gen$, $\AA_n$ a sequence of asset sets, $\asset = (\asset_n)$ an asset and $\verify$ an efficiently computable asset verifier.
A \textbf{secure obfuscator} for $(\C, \AA, \asset)$ is a randomised algorithm $\OO(P, \aux)$ that transforms a program $(P,\aux) \leftarrow \Gen_\C(n)$ in $\C_n$ to a program $P' = \OO(P,\aux)$ in $\C_n'$ for some class of programs $\C'= ( \C_n' )$.
Further, we require:
\begin{enumerate}
\item (Correctness) For all $P \in \C$ and all\footnote{One can relax this to almost all inputs if necessary.} inputs $x$ we have $P'(x) = P(x)$.

\item (Efficiency) There is a polynomial $p_1(t) \in \R[t]$ such that $| P' | \le p_1( |P| )$ for all $P \in \C$ (for some canonical measure $|P|$ and $|P'|$ of program size in the languages of $\C$ and $\C'$).
There is a polynomial $p_2(t) \in \R[t]$ such that if $P(x)$ runs in time $T$ for an input $x$ then $P'(x)$ runs in time $p_2(T)$.

\item (Security)
For all probabilistic polynomial-time algorithms $A$ that know $\Gen_\C, \OO, \verify$, such that $A : \C_n' \to \AA_n$ the probability 
\[
    p_n = \Pr\Bigl( \verify( P, \aux, A( P' )) = 1 :  
\]
\[  (P,\aux) \leftarrow \Gen_\C(1^n) , P' = \OO( P, \aux ) \Bigr) \]
over $( P, \aux) \leftarrow \Gen_\C$ and the random choices of $\OO$ and $A$,
is negligible in $n$.
By $p_n$ being negligible in $n$ we mean: for all polynomials $q(x) \in \Z[x]$ there is some $N$ such that, for all $n > N$, $p_n \le 1/|q(n)|$.
\end{enumerate}
\end{definition}

One of the main motivations behind our formalism is to be able to write down the 
equation in item (3) of the definition,
which gives falsifiable criteria for security of an obfuscator.
Formulating this security property requires the ability to sample $P \in \C_n$ and the ability to determine whether or not an adversary has succeeded.
Note that it is not necessary that $\asset_n( P )$ is efficiently computable, but it is necessary that an asset is efficiently verifiable.
The connection between this definition and setting up an obfuscation challenge is given in Section~\ref{sec:obfuscation-challenges}.

Note that:
\begin{enumerate}
\item We do not specify the power of the attacker $A$, apart from that it is provided with the obfuscated program $P' = \OO( P, \aux ) $.
At a purely formal level, there is no distinction between an attacker who inspects the source of the program, or who runs a symbolic analyser on the program, or who executes the program in a debugger.
Hence our formalism includes the strongest possible dynamic attacker.

\item Our notion is falsifiable, meaning that if one makes a claim that an obfuscator protects a given asset in a given class of programs, then a cryptanalyst can give an explicit counterexample to show the claim is false.
An objection to some previous work on obfuscation is that the definitions are so vague it is not even clear what it means to show an attack or counterexample.
For further discussion, see Section~\ref{sec:obfuscation-challenges}.


\item A natural question is how to combine obfuscators that are designed to protect different assets, in order to get one program with all assets protected.
Section 2.1 of Schrittwieser \etal \cite{SKKMW16} mentions ``combinations of different types of obfuscators'' and states that ``a systematic analysis of how obfuscation techniques compose with one another would be challenging'' and that it ``makes an interesting and relevant topic for future work''.
We discuss this problem and give new results in Section \ref{sec:composition}.

Be aware that the phrase ``composing obfuscators'' exists in the theoretical literature, but means something different.


\end{enumerate}

Finally, Garg and Pandey~\cite{GP17} have considered program obfuscation under updates.
They note that a small incremental update to the source program should only require a small update to the obfuscated program. 
Our approach to special-purpose obfuscation naturally achieves this goal.

\section{Connection with Obfuscation Challenges} \label{sec:obfuscation-challenges}

It may be helpful to the reader to see that the problem of defining obfuscation security is exactly the same as providing automated obfuscation challenges.
For example, the Tigress project~\cite{tigress,banescu2016code,BCP17}
provides reverse engineering challenges that are obfuscated programs with a cash prize for the first contestant to successfully reverse-engineer them.
Such competitions are a popular way to encourage research into breaking obfuscation schemes, and to give insight into the security of current obfuscation methods.
To set up such a competition, one needs a means to generate random programs to be obfuscated, and a means to decide whether a contestant has successfully completed the task.

Banescu, Collberg, and Pretschner~\cite{BCP17} build a ``code generator that can generate large numbers of arbitrarily complex random C functions''. 
Essentially, their generator produces a random hash function of a certain type, and the program is a licence check. 
Determining the success of the contestant is clear, \ie find an input that is accepted by the licence check.
This generator has been implemented in Tigress as \texttt{RandomFuns}.

One sees that the requirements to set up a successful obfuscation challenge are exactly captured by our formalism. 
First, one needs to have a well-defined class of programs and an algorithm $\Gen$ to generate a random program from this class. 
Second, one needs a well-defined security goal (the asset).
Third, one needs an efficient and automated method to judge if a contestant has found the asset (the asset verifier). 
A potential future application of our formalism is to set up obfuscation challenges for programs other than simple licence checks.

The issue of learnability also arises in this context.
If the asset is guessable, or learnable from the input-output behaviour, then an adversary can win the game without reverse-engineering the code.
Hence, it only makes sense to offer an obfuscation challenge when the asset is not learnable.
This requirement is implicit in the previous work.

Finally, there is an interesting subtlety about determining whether a competitor has successfully determined the asset.
In our formalism, the verification algorithm $\verify$ takes inputs $( P, \aux,  a )$, and so requires the original program and the auxiliary data.
This seems to be necessary in some contexts; whereas, in other contexts (\eg the ones studied in~\cite{BCP17}), one can verify the asset using only $P' = \OO( P, \aux ) $.
It follows that obfuscation challenges may come in two flavours: those that are publicly verifiable, and those for which only the problem-setter can decide if a solution is correct.\footnote{Similar issues arise in some other crypto challenges, such as LWE challenges.}


\section{Formalism in Action}
\label{sec:act}
The purpose of this section is to illustrate how our formalism is intended to be used. 
We do this using a very simple example, namely \emph{point functions}, which include licence checks and password checks.
Point functions may seem a very primitive class of programs, but they have applications in practice.
For example, Sharif \etal \cite{SLGL08} consider a predicate that triggers malicious behaviour in malware and encrypts the program block using the satisfying input as the key.

Obfuscating point functions is already well-understood.
The goal of this section is simply to give a clear explanation of how to use our formalism in practice.
First, we show how to define security of a point function obfuscator using our formalism, and argue that this agrees with an existing notion in the literature.
Second, we show a good point function obfuscator, using our definitions.
In Section~\ref{app1}, we show how to disprove security of a bad point function obfuscator, using our definitions.

\subsection{Defining Security of Point Function Obfuscation}\label{sec:point-functions}

The class of point functions is defined to be the set of program segments (or functions) that take an input $x$ and output $1$ (or execute a certain code segment) if and only if $x$ is equal to some particular value $c$.
The classic example is a licence check or password check.

The first step of putting this into our formalism is to define the class of programs.

\begin{definition}
Let $\C_n$ be the set of programs that take an $n$-bit input $x$, and are zero on all except one input $c \in \{ 0,1 \}^n$ (the password or the licence code). 
So, $P(x)$ is the predicate \texttt{x == c}.
We call $\C_n$ the class of point function programs.
\end{definition}

Each choice of $c \in \{ 0,1 \}^n$ defines a different point function in $\C_n$, so $| \C_n | = 2^n$.
The crucial fact that makes this problem interesting is that when $n$ is large then an attacker cannot guess $c$ with non-negligible probability.

The next step of putting this in our formalism is to define the program generator $\Gen_\C$.

\begin{definition}
The program generator $\Gen_\C$ for point functions takes input $n$ (technically speaking, the input should be written in unary as $n$ ones), samples uniformly at random a binary string $c \in \{0,1\}^n$, sets $\aux = c$, and returns a program $P(x)$ given by

\begin{center} \verb+if (x==c) then return 1 else return 0+ \end{center}

\end{definition}


Now, we define the assets in this program class. 
The asset in this case is the secret input $c$. 
So, we define $\AA_n = \{ 0, 1\}^n$ and $\asset_n : \C_n \to \AA_n$ so that $\asset_n(P) = c$ is the specific value such that $P( \asset_n(P)) = 1$.

The formalism also requires us to specify an asset verifier. 
To verify that an adversary has computed the asset of a program correctly, we simply define $\verify(P, \aux,  \asset_n(P))$ to check that $\aux = \asset_n(P)$.
Alternatively, in this case, the auxiliary information is not required and we can verify using $P$ alone by computing $P( \asset_n(P)) \in \{ 0,1 \}$; since $P$ is computable in polynomial time, $\verify$ is computable in polynomial time.

An \emph{obfuscator for point functions} is a program $\OO$ that takes as input a program $P$ from the class $\C_n$ and outputs a program $P'$ in some other class $\C'_n$. 
The class $\C'$ depends on the details of the obfuscator; we will see an example in the next subsection.
As in Definition~\ref{defn:secure-obfuscator}, we require $P'$ to be correct (having the same functionality as $P$) and be efficient to execute.

The final part of our formalism is to define security of an obfuscator $\OO$.
Consider a probabilistic polynomial-time algorithm $A$ that attacks the system. 
The attacker $A$ has full knowledge of $\Gen_\C, \OO$ and  $\verify$, which is consistent with Kerckhoff's principle.
The attacker $A$ takes as input $P' \in \C'_n$ and outputs an element of the asset space $\AA_n$, namely  a value $c \in \{0,1\}^n$. 
The attacker wins if $\verify( P, \aux, A( P' )) = 1$, in other words if $c$ is the accepted input by the point function.
We say that the obfuscator is \emph{insecure} if the probability $A$ wins, over uniformly chosen $P \in \C_n$ and over the random choices of $\OO$ and $A$, is noticeable.
More precisely, we define $ p_n = \Pr\Bigl( \verify( P, \aux, A( P' )) = 1 \Bigr)$ for each $n$, and say that $A$ breaks the obfuscation scheme if $p_n > 1/|q(n)|$ for some polynomial $q$ and all large enough $n$.
Conversely, we say that the obfuscator is \emph{secure} if, for all probabilistic polynomial-time algorithms $A$, the success probability $p_n$ of $A$ is negligible.

In 2014, Barak \etal \cite{BBCKPS14} defined \emph{input-hiding security}: It should be hard for an algorithm $A$, given $P'$ for $P \in \C_n$, to find an $x \in \{0,1\}^n$, such that $P( x) = 1$.
Therefore, the security definition above, defining the asset $\asset_n(P)=c$, is equivalent to input-hiding security.
This shows that our formalism is powerful enough to include existing security definitions from theoretical cryptography.

We also remark that a more general formulation of this problem, as ``Constant Hiding'' is given by Kuzurin \etal~\cite{KSVZ07}.

\subsection{A Secure Obfuscator for Point Functions}

Obfuscating point functions is well-known.
It received its first theoretical treatment by Canetti~\cite{Can97}.

The classic obfuscator $\OO$ for point functions is to use a cryptographic hash function $H$ with $n$-bit output.
If the program $P \in \C_n$ is

\begin{center} \verb+if (x==c) then return 1 else return 0+ \end{center}

\noindent then the obfuscated program $P'(x)$ includes constants $r$ and $c' = H( r \Vert c )$ and checks whether or not $H( r \Vert x ) = c'$.

In other words, $\OO$ is a randomised algorithm that takes $P \in \C_n$ as input, determines $c$ from examining the code or from the auxiliary input $\aux$, chooses a random $r \in \{0,1\}^n$, sets $c' = H(r \Vert c )$ and outputs the program $P'(x)$ given by

\begin{center}
 \verb+if (H(r || x)==c') then return 1 else return 0+ .
\end{center}

Lynn, Prabhakaran, and Sahai~\cite{LPS04} prove security of obfuscating point functions in the random oracle model, and the problem has also been considered by Hofheinz, Malone-Lee, and Stam~\cite{HMLS07}
and Di Crescenzo~\cite{DC18}.
In particular, Section~3 of Di Crescenzo~\cite{DC18} gives the same construction as we give.

%

We now show how to express such a security proof in our formalism.
With our formulation, random oracles are not required and we can give a proof in the standard model based on a fairly standard assumption about hash functions.

\begin{definition} \label{defn:hash-inversion}
(Hash inversion problem)
Let $\{ H_n \}$ be a family of hash functions $H_n : \{ 0,1 \}^* \to \{0,1\}^n$.
We call $\{ H_n \}$ \emph{smooth} if, given uniformly sampled $r, y \in \{0,1\}^n$, the probability that there exists an $x \in \{0,1\}^n$ such that $H_n( r \Vert x)  = y$ is at least $1/n$.

The \emph{hash inversion problem} is: Given $n \in \N$ and $y_n \in \{0,1\}^n$, to compute $x \in \{ 0,1 \}^*$ (if it exists) such that $H_n(x) = y_n$.

The \emph{hash inversion assumption} for a family of hash functions is that if $y_n$ is sampled uniformly at random then it is hard to compute $x \in \{ 0,1 \}^*$ such that $H_n( x ) = y_n$.
More precisely, the hash inversion assumption is that for all probabilistic polynomial-time algorithms $A$, the probability (over choices for $y_n$), that $A( n, y_n ) = x \in \{0,1\}^*$ such that $H_n(x) = y$, is negligible as a function of $n$.
\end{definition}

\begin{theorem}
The point function obfuscator implemented with a smooth hash family $\{ H_n \}$ is secure 
in the standard model if the hash inversion problem holds for the family $\{ H_n \}$.
\end{theorem}

\begin{proof}
Let $(n, y_n \in \{0,1\}^n )$ be a hash inversion challenge for the hash function $H$.
Let $A$ be an attacker for the point function obfuscator; so $A$ takes as input an obfuscated program $P'$ and outputs a binary string $x$ that is accepted by the program.

Choose a random binary string $r \in \{0,1\}^n$ and set $P'$ to be the obfuscated point function program (here $H = H_n$ and $y=y_n$)

\begin{center} \verb+ if (H(r || x)==y) then return 1 else return 0.+ \end{center}

\noindent
Since $y_n$ and $r$ are sampled uniformly and $H_n$ is smooth, there is a noticeable probability that there is some $x \in \{ 0, 1\}^n$ such that $H_n( r \Vert x ) = y_n$, in which case $P'$ is the obfuscation of a randomly sampled program in $\C_n$ (it just so happens we don't know $P$).

Now, run the adversary $A$ on $P'$. 
The adversary outputs an asset $x \in \{ 0, 1\}^n$ and wins if $P'(x) = 1$.
But, this means that $H_n(r \Vert x)  = y_n$ and so we return the binary string $r \Vert x$ as the solution to the hash inversion challenge.
This completes the proof. \qed
\end{proof}


To conclude, we have shown in this section how to define an obfuscator using our formalism, and how to prove security.

\subsection{Insecure obfuscator for point functions}\label{app1}

The purpose of this section is to show how our formalism is falsifiable, in the sense that it is explicit what it means to break an obfuscator. 
This ties in with the discussion in Section~\ref{sec:obfuscation-challenges}.

So, consider the following (obviously trivial) obfuscator $\OO$ for point functions: The input $P(x) \in \C_n$ to $\OO$ is a program 

\begin{center} \verb+if (x==c) then return 1 else return 0+ . \end{center}

\noindent Then, $\OO$ determines $c$ from examining the code, chooses a random $r \in \{0,1\}^n$, sets $c' = r \oplus c $, and outputs the program $P'(x)$ defined by

\begin{center}
   \verb+if ((r^x)==c') then return 1 else return 0+ .
\end{center}

To show that, this obfuscator is insecure we are required to define an attacker $A$. 
The attacker $A$ takes as input the source code of $P'$. 
Remember that the algorithm $A$ knows the class $\C_n$ and knows exactly what $\OO$ is doing.
Hence, $A$ simply identifies the values $r$ and $c'$ from the code, computes $x = r \oplus c'$, and returns $x$. 
It follows that $A$ succeeds with probability $1$, since $x$ is exactly the asset in the program.

\section{ Real-World Examples} \label{sec:examples}   

We now consider some classes of programs that are natural targets for obfuscation in real-world applications.
Our goal is to show how the security of obfuscation in these cases can be expressed using our formalism. 
This will demonstrate how our formalism can be used to capture security goals in a wide variety of contexts.
The aim of this section is not to give complete obfuscation solutions; however, we do give some sketches and references for how to solve these problems.

\subsection{Evasive functions} \label{sec:evasive}

A class of functions $\C_n = \{ f : \{0,1\}^n \to \{0,1\} \}$ is \emph{evasive} if for all $n$ and for all $x \in \{0,1\}^n$ and all $f \in \C_n$ the probability that $f(x) = 1$ is negligible (with respect to $n$).
We call an $x \in \{0,1\}^n$ such that $f(x) = 1$ an accepting input.
Point functions are evasive, but there are also many other classes such as: hyperplane membership~\cite{CRV10}, logic formula defined by many conjunctions~\cite{BR17}, pattern matching with wild cards~\cite{BKMPRS18}, root of a polynomial system of low degree~\cite{BBCKPS14}, compute-and-compare programs~\cite{WZ17,GKW17}, and more~\cite{LPS04}.
Each of these classes of programs can be described by a program generator $\Gen$.

A nice example of evasive functions is a software patch for a rare bug -- this is explained in~\cite{BBCKPS14} (they credit Ganesh, Carbin, and Rinard~\cite{GCR12}).

It is important that satisfying inputs cannot be guessed. 
The word ``evasive'' is evocative of satisfying inputs being rare and hard to find.

Barak \etal~\cite{BBCKPS14} define \emph{input-hiding obfuscation} for evasive functions: It should be hard for an algorithm $A$, given $\OO( P )$ for $P \in \C_n$, to find an $x \in \{0,1\}^n$, such that $P( x) = 1$.
We now explain how to express their definition in our formalism. 
Let $\AA = \{0,1\}^n$ and $\asset( P ) = \{ x \in \{0,1\}^n : P(x) = 1 \}$.
Note that $\asset(P)$ is not an element of $\AA$, but is a subset of $\AA$.
One issue is that the size of $\asset(P)$ may be quite large, and it may be computationally difficult to compute all of them.
Indeed, the only condition required for being evasive is that $\#\asset(P) / 2^n$ is negligible, and so $\#\asset(P)$ may be exponentially large (in $n$).
Hence, it does not make sense to define an attacker to be an efficient algorithm that outputs the whole set $\asset(P)$.
Instead, we consider an adversary who takes $P$ as input and returns $a \in \{ 0, 1 \}^n$.
We now define an efficiently computable asset verifier for this problem: On input $(P,a)$ the algorithm $\verify$ runs $P(a)$ and returns the output of $P(a)$.
In other words, $\verify(P, a) = 1$ if and only if $P(a) = 1$, and since $P$ is polynomial-time, we have that $\verify$ is also polynomial-time.

In summary, our formalism can also include the general notion of input-hiding for evasive functions.

Barak \etal~\cite{BBCKPS14} make some interesting remarks about the security of evasive functions.
They prove that input-hiding obfuscation is not equivalent to VBB obfuscation in general (they give a trivial separating counterexample in Section 2.2.1).
Indeed, they show the notions are \emph{incomparable} (VBB does not imply input-hiding; input-hiding does not imply VBB).
On the other hand, they prove that the two notions are equivalent in the case of ``polynomially-sparse evasive collections'', which are function classes such that the sets $\{ x \in \{0,1\}^n : f(x) = 1 \}$ are all polynomially sized (in $n$).
Of course, a polynomially-sized evasive class can be reduced to just executing polynomially many point functions, so it is not very interesting.
These remarks support our claim that, not only is VBB obfuscation too strong for most practical situations, it is not even necessarily the desired security property.
Instead, we believe that formalising security in terms of assets is clearer and more fine-grained.

Finally, we remark that the class $\C$ of all possible evasive functions on $n$-bit inputs is probably too general a class to be considered.
In practice, one will consider specific classes, such as $\C$ being the set of all hyperplane membership functions~\cite{CRV10}, or the set of all pattern-matching-with-wildcard functions~\cite{BKMPRS18}. 
Having a more restricted class allows to specify $\Gen_\C$ more easily, and also allows to consider more efficient special-purpose obfuscators.

\subsection{Biometric Matching}

When a user provides a biometric reading (\eg fingerprint or facial image) then certain features are extracted and represented as a \emph{template} $t$.
When a user submits their biometric reading then there may be measurement errors. 
The process of determining if the biometric matches the template is called \emph{fuzzy matching}. 

Most biometric-based authentication systems provide some privacy of the template biometric data, by storing some kind of \emph{tag} $t'$ that is like an encrypted or obfuscated version of the template. 
When a user submits their biometric reading then the process of determining if the biometric matches the template can be viewed as an \emph{obfuscated fuzzy matching} program. 
There are a number of solutions in the literature~\cite{TAKSBV05}.


We show how to express security in our formalism.

\begin{definition}
For $n \in \N$, consider the set of all $n$-bit binary strings (biometric readings) and let $T_n$ be the set of all possible templates $t$ extracted from them.
The set $\C_n$ of programs is parameterised by $T_n$. 
For each template $t \in T_n$, we have a program $P \in \C_n$ that takes as input an $n$-bit binary string $x$ and outputs $1$ if and only if the extracted template from $x$ matches $t$ within the given error tolerance.

The algorithm $\Gen_\C$ selects uniformly at random an $n$-bit string and outputs this program $P$ together with $\aux = t$.

The asset $\asset_n( P )$ is the template $t \in T_n$. 
So, $T_n$ is the asset space.

An obfuscator $\Obf$ for fuzzy matching is secure if no attacker can efficiently compute the asset $t$ when given an obfuscated program $P'(x)$.
\end{definition}

The common security model for secure biometric matching is actually information-theoretic: Given a tag $t'$ that is to be matched by an input, there is still substantial entropy in the possible template $t$.
Solutions to the biometric problem are often based on concepts like \emph{secure sketches} and use coding-theoretic ideas.
Instead, our formalism gives an alternative way to define security in the context of fuzzy matching, which is more amenable to solutions based on \emph{computational security} than information-theoretic security.
Obfuscation solutions in our model have been given by~\cite{KC16,GZ18}.


\subsection{White-Box Cryptography and Digital Rights Management (DRM)}\label{sec:WBC}

White-box cryptography was introduced by Chow \etal \cite{CEJO02,CEJO02a} for digital rights applications.
DRM is an access control mechanism that aims at restricting access to digital contents. 
In the context of obfuscation, white-box cryptography and DRM are discussed in~\cite{Joye08,Sethi-MSc04-DRM-Obf}

The most common scenario in white-box cryptography is a decryption algorithm for a known symmetric cipher with key $k$.
The idea is that one can provide a user with such a program, and the user cannot determine the key.
This is one of the standard security models for white-box cryptography, and we now show how it can be expressed using our formalism.

\begin{definition}
Let $Enc$ and $Dec$ be the encryption and decryption algorithms for a block cipher  with variable key length.
Define $\C_n$ to be the set of programs $P = Dec_k$ for each possible $n$-bit key $k$.

Choosing $P \in \C_n$ is simply choosing a secret key $k$. 
So, the algorithm $\Gen_{\C}(1^n)$ samples an $n$-bit key $k$ and outputs $P = Dec_k$ and $\aux = k$.

Define $\asset(P) = k$ to be the secret key corresponding to that instance.

For a candidate asset $a$, the asset verifier $\verify( P, \aux, a)$ just checks whether $a = \aux$.

An obfuscator $\Obf$ for white-box encryption is secure if no attacker can efficiently compute the key $k$ when given an obfuscated decryption program.
\end{definition}

An alternative verify algorithm (which is probabilistic) is to check the correctness of the key by encrypting random messages using the candidate key and seeing if the white-box program decrypts to the correct message.

A wider range of security models for white-box cryptography are given by Delerabl{\' e}e \etal in~\cite{DLPR14}, Saxena \etal \cite{SWP09} and Bogdanov and Isobe \cite{BI15}.\footnote{Note that Delerabl{\' e}e \etal \cite{DLPR14} also consider active attack models where the attacker is given more than just the obfuscated program. 
We do not consider such models in this paper.}
The basic notion already discussed is called \emph{unbreakability} in~\cite{DLPR14}, but there are other notions such as \emph{one-wayness} (given an obfuscated program to compute $Dec_k$, it should be hard to make a program to compute $Enc_k$) and \emph{incompressibility} (hard to write a shorter and/or more efficient program with the same functionality).
These other notions can also be expressed in our formalism, as we now very briefly explain.
For one-wayness, we define the asset to be any program that computes $Enc_k$; for incompressibility, we define the asset to be any program that computes $Dec_k$ correctly and whose size is below some threshold.
In both cases, the verifier can work by choosing some random messages/ciphertexts and testing the outputs, in which case the verifier is randomised and only gives a correct result with some probability.

Further discussion of obfuscation in the context of DRM is given in~\cite{AhPrKe2018}.

\draft{
A slightly related topic is studied by Hohenberger \etal \cite{HRSV11}. 
They give a method for obfuscating re-encryption.
Re-encryption means transforming a ciphertext for a message $m$ encrypted under Alice's public key into a ciphertext for the same message $m$ under Bob's public key.
The asset can be either Alice's or Bob's private key.
}

\subsection{Finite Automata}\label{sec5.5}

A finite automaton (state machine) is a program to recognise whether or not an input $x$ is a member of a language $L$. 
It has a set of states and, for each symbol in the alphabet, a ``next state'' transition matrix. 
One can imagine a scenario where one wants to compute the automaton but keep the transition matrix secret. 
A related application is regular expressions.
The question of obfuscating automata is mentioned in Section 6 of Lynn, Prabhakaran and Sahai~\cite{LPS04} and in Kuzurin \etal~\cite{KSVZ07}.

State machines are used in many applications, such as Cyber-Physical Systems (CPS) and control systems (\eg Supervisory Control And Data Acquisition -- SCADA in short), to encode workflows and other processes in industrial and operational systems \cite{MFXLL15,IMLFL16}.
These processes may be finely tuned and optimised using complex mathematical models, or as the result of years of experience.
It may arise that these workflows need to be protected against tampering, or that we want to prevent leaking the workflow information to a competitor.
Hence, an organisation may wish to prevent theft of the intellectual property represented in such a process, yet the state machine/process itself may be implemented in software throughout the organisation using embedded devices or Internet-of-Things (IoT)-enabled devices.
Moctezuma \etal \cite{MFXLL15} discuss the role of state machines in CPS while Iarovyi \etal \cite{IMLFL16} describe the security concerns in this context.

We now explain how to express security using our formalism.

\begin{definition}\label{defn:automata-class}
Define $\C_n$ to be the set of state machines with $n$ states and alphabet $\{0,1\}$. 
Let state $1$ be the initial state and state $n$ be the accepting state.

Each element $P \in \C_n$ recognises some language $L \subseteq \{0,1 \}^*$.
One can generate a random program in $\C_n$ by choosing randomly a pair of transition matrices $(M_0, M_1)$; hence, we have a program generator $\Gen_\C$ and the value $\aux$ is the pair $(M_0, M_1)$.
\end{definition}

A natural definition for the asset $\asset(P)$ is the pair of transition matrices, but note that this depends on the exact ordering of the states; whereas, we might consider two state machines to be isomorphic if each transition matrix of one of them is the conjugate by a permutation matrix of the corresponding transition matrix of the other.
Even more problematic is the fact that there may be many quite different state machines that recognise the same language.
Hence, we need to be more flexible in our definition of the asset.

Our formalism is designed to handle such issues. 
In this case, $\asset(P)$ is \emph{any} pair of transition matrices for an automaton that recognises the same language. 
We then require the asset verifier to determine if two state machines recognise the same language.
Fortunately, it is known that one can efficiently test equivalence of automata/regular languages (see Hopcroft and Karp~\cite{HK71} or Almeida \etal \cite{AMR10}).

\begin{definition}
Let notation be as in Definition~\ref{defn:automata-class}. 
Then, $\asset(P)$ is any transition matrix pair for an automaton that recognises the same language as $P$.
The asset space for $\C_n$ is the set of all pairs of well-formed $n \times n$ transition matrices.
For a candidate asset $a$, the asset verifier $\verify( P, \aux, a)$ is an efficient algorithm to determine if the two automata (given by transition matrix pairs $\aux$ and $a$ respectively) recognise the same language
\end{definition}

Finally, to ensure that security is possible, we need to consider whether a finite state machine is learnable from its input/output behaviour.
Angluin~\cite{Ang87} shows that one can learn an automaton when given a way to generate random inputs that are accepted and also random inputs that are not accepted. 
However, if it is hard to efficiently find such inputs from ``black box'' execution of the machine, then it seems to be hard to determine the automaton. 
Hence, the obfuscation problem makes sense for sufficiently complex finite state machines for which either the language or its complement is small.

We therefore need to restrict to the subclass of $\C_n$ of evasive automata.
One then can consider an obfuscator $\Obf$ that takes as input a transition matrix pair for an automaton $P$ and outputs an obfuscated program $P'(x)$ that takes $x \in \{0,1\}^*$ as input and returns $1$ if and only if $x$ is in the language of $P$.
The obfuscator is secure if no adversary takes $P'$ as input, and outputs a transition matrix pair for an automaton $P$ that recognises the same language.
Galbraith and Zobernig~\cite{GaZo20} have proposed an obfuscation scheme for automata that satisfies this definition.

Definition 3 of Kuzurin \etal~\cite{KSVZ07} introduces the notion of a ``software protecting obfuscator'', which is similar in flavour to iO, and proposes this notion as suitable for the problem of obfuscation DFA (Deterministic Finite State Automata). 
We believe circuit-hiding and/or input-hiding are more natural security definition for this problem.

\subsection{Machine Learning and Artificial Intelligence (AI)}\label{sec5.6}

Many current applications of machine learning (such as drug design in the pharmaceutical domain \cite{Burbidge-CC01-SVM-Drug} and Industrial Control Systems (ICS) \cite{Fukuda-ToIE92-ICS}) require large proprietary datasets and massive computational investment in the training stage. 
But once training is completed, the actual program may be relatively compact.
The compact program can be viewed as a ``representation'' of the knowledge extracted from the dataset.\footnote{Whether this knowledge representation is useful is an interesting question. 
Indeed, a neural net may be viewed itself as a form of obfuscated program.}
An adversary who does not have access to the raw data, or who does not have the resources to perform the training, may potentially be able to extract sensitive knowledge from the source of the program. 
So, if the AI program becomes a component of a product whose code is available to an adversary (\eg a mobile app) then one might be concerned to protect the intellectual property embedded in the program.

Song \etal~\cite{SRS17} give attacks to extract private information from the training data from the final machine learning model. 
They discuss attacks in the \emph{white-box case} (where the attacker can access the model directly) and the \emph{black-box case} (where the attack only has input/output access to the model).
One could use obfuscation to prevent such white-box attacks, though this is only interesting when the data remains secure against a black-box attack.

The majority of pattern matching and neural net algorithms contain a small set of basic functions. 
The basic operations include convolutions, linear or affine maps on $\R^n$ for large $n$ and activation functions.
The basic topology of the program (order of operations, data flow) may be public.
But the main items that require protection include, but are not limited to, the weights, parameters, and thresholds. 
Typically, these appear in the program as constants.

Therefore, we consider a class $\C$ of programs that perform sequences of these operations on large dimensional inputs.
The asset is the list of weights and thresholds.
An obfuscator should allow to compute the output of the neural net/AI, without revealing the data in the model/net.
The algorithm $\verify$ checks equality of constants (possibly up to some rounding accuracy).


\draft{
\subsection{Complicated Mathematical Functions or Processing}\label{sec5.8}

There are situations where a program segment computes some mathematical transformation or processing whose formula constitutes intellectual property.
We list some examples of intellectual property in mathematical computation, and where program obfuscation is a natural solution.
%
\begin{enumerate}
\item Scientific computing or mathematical software.
Many of the leading scientific computing packages are proprietary (\eg Mathematica, MATLAB).
Particular algorithms may have intellectual property that is worth protecting, such as efficient algorithms for solving certain numerical problems or optimisation problems.

\item File compression software may use a particular algorithm or be proprietary to enable DRM features.

\item Software tools for analysis of ``big data'' may contain proprietary algorithms~\cite{Bur16,Dia15}.

For example, consider feature recognition from complex data, such as a program that recognises people from a video feed.
If the pattern recognition code is proprietary then we may wish to obfuscate the code to prevent others from being able to deploy the pattern recognition code in their own application.

Proprietary algorithms may also be used in libraries to create neural networks: These algorithms can be trained on user-provided data and then provide a predictive model for future use by the user.
The precise learning algorithms used in some of these packages are intellectual property of the software company and are required to be protected.

\item Software agents for online auction sniping~\cite{TRLA11} may contain decision-making strategies that, if known to a competitor, could be circumvented.

\item A chess program may contain proprietary algorithms that are the result of significant software development, and the designer may wish to protect them from theft.

\item A financial services company may develop and sell proprietary market prediction software (\eg see~\cite{BMZ11}). 
The company may wish to keep the precise statistical tools and formulae secret. 
\end{enumerate}
%
%
Hence, it might be required to obfuscate the computation of these functions, to make it hard for an adversary to understand what processing is being performed.

Of course, there is no security benefit to obfuscate the computation of a mathematical function if that function itself can be efficiently determined by its behaviour on inputs. 
For example, low degree polynomial functions (even in several variables) can be interpolated from random input/output pairs.
So, we need to focus on classes of functions that are not learnable.

Zhou \etal~\cite{ZMGJ07} have considered obfuscating the computation of functions comprised of basic boolean and integer operations.
They do this  by finding an equivalent, but more complex, formula that computes the same function.
Eyrolles \etal~\cite{EGV16} have cryptanalysed their work.

We show how to express this situation in our formalism.
Suppose $\C$ is a class of programs (or program segments) that contains some complicated mathematical processing $f : \R \to \R$ or $f : \{0,1\}^n \to \{0,1\}^n$ formed by combining various standard mathematical functions.
We assume that functions in $\C_n$ are ``elegant'' in some sense (that depends on the context of the application), which means that should have a relatively short representation (\eg power series with a small number of terms; linear combinations and compositions of functions in some library; \etc).
Specifying the precise meaning of ``elegance'' is equivalent to explaining how to construct a program generator $\Gen_\C$ to output a random function in the class $\C_n$.
The intellectual property is the elegant expression that computes some useful transformation or mathematical operation. 

We define the  asset $\asset(P)$ to be a formula that satisfies this notion of elegance.
The asset verifier is required to check that $\asset(P)$ is short and simple, and also to check that it computes the same function as the program $P$.
This latter requirement may be hard to satisfy for some program classes, but one can always build a randomised verifier that just chooses some random inputs $x$ and checks whether it is always the case that $\asset(P)(x) = P(x)$
(or, for example when using floating-point arithmetic, that $| \asset(P)(x) - P(x)| < \epsilon |P(x)|$ for some small $\epsilon  > 0$).


%
%

%
}

\draft{
\subsection{Exception Handling, Errors, and Bugs}\label{sec5.9}

Handling of invalid inputs, special cases, numerical errors, overflows, and bugs is essential part of almost every program~\cite{Wirfs-Brock-Soft06-Exceptions}.
Sometimes, a developer might wish to hide these special events, perhaps to prevent an adversary from deliberately making the program enter an error state, or because there is intellectual property that is revealed when handling special cases.
To achieve such protection, a software developer could obfuscate the conditional branch and/or code segment where errors and bugs are handled. 
The adversary will win if it can identify inputs that satisfy the condition and/or the code block that handles the exception.

One can formulate this within our model. 
For a class $\C$ of programs, we define the asset $\asset$ to be the set of conditional expressions that correspond to error events, or the set of program blocks executed during error handling.

%

A related use of obfuscation is to make it hard for an attacker to find vulnerabilities and flaws in programs.
There are many examples of exploits and vulnerabilities caused by software bugs like lack of input validation, buffer/stack overflow, and string format vulnerability.
Routinely obfuscating programs would make it harder for attackers to find such vulnerabilities.
However, this scenario cannot be captured within our formalism.
Our formalism requires the developer to have a precise and well-defined notion of asset; whereas, this scenario is about ``unknown unknowns'': potential vulnerabilities whose form and existence are not known to the developer (if they were known then the developer could guard against them).
}

\draft{
\subsection{Watermarking}

Watermarks are introduced into code so that copying can be detected and the source of the piracy traced.
In this case, the asset is part of the obfuscated program, and not the original program, so it is not quite the same as in our formalism. 
Nevertheless, we can use our formulation to express the success of the adversary.

The main adversarial goal in watermarking is to locate and remove/modify the watermark, we refer to Myles and Collberg~\cite{MC06} for details.
Hence, an asset in an obfuscated program is the location of the watermark. 
The algorithm $\verify$  just determines whether the adversary has correctly identified the location of the watermark.
So, the security of watermarking can be formulated in our formalism.
}

%

%
%

%


\section{Composition of Obfuscations} \label{sec:composition}

In this section, we discuss the security of applying different obfuscators iteratively on a program, to protect multiple assets of different types.
We call this \emph{composition} of obfuscators.
This concept has been discussed by researchers in the software security world and seems to be commonly used.
For example, Section 2.1 of Schrittwieser \etal \cite{SKKMW16} note that ``many commercial obfuscators employ (and indeed recommend to use) multiple obfuscations at the same time''.
They describe this as an important topic for future work.
But composition of obfuscators has never had a rigorous theoretical treatment.

Be aware that the word ``composition'' already has several meanings in the context of obfuscation.
For example, Lynn, Prabhakaran, and Sahai~\cite{LPS04} use the phrase ``compositions of obfuscations'' to mean  applying the same obfuscator $\Obf$ to a sequence $P_1, \dots, P_t$ of programs to get $(\OO(P_1), \dots, \OO( P_t ))$.
They write ``we use the term compose in the same way as one refers to composition of cryptographic protocols -- to ask whether having multiple instances in the system breaks the security or not''.
Hofheinz, Malone-Lee, and Stam~\cite{HMLS07} also discuss composability in the context of point function obfuscation: ``security may be lost when the obfuscation is used in larger settings in which $x$ is used in several places''.

We stress that our interest is in a class of programs $\C_n$ that may have several different assets, and where obfuscators $\Obf_1$ and $\Obf_2$ are each designed to protect a different asset in a different way.
We wish to have an assurance that the composition $\Obf_2( \Obf_1( P ))$ protects both assets (a formal description is given in Theorem~\ref{thm:composition}). 
This topic is un-interesting when discussing VBB or iO obfuscation, since one pass by such an obfuscator already hides all information about $P$ that is not learnable from input/output behaviour.
This is one reason why the theoretical community uses the word ``composition'' differently.

The following theorem gives some conditions under which composition of obfuscators makes sense and maintains security.

\begin{theorem}\label{thm:composition}
Let $\C_n$ be a class of program segments defined by a generator $\Gen_\C$.
Let $\asset^{(1)} : \C_n \to \PP( \AA_n^{(1)} )$ and $\asset^{(2)} : \C_n \to \PP( \AA_n^{(2)} )$ be assets with verifiers $\verify^{(1)}$ and $\verify^{(2)}$.
Suppose we have a secure obfuscator $\Obf^{(1)}$ that protects $\asset^{(1)}$ and a secure obfuscator $\Obf^{(2)}$ that protects $\asset^{(2)}$.
Suppose further that the following conditions hold.
\begin{enumerate}
\item $\Obf^{(1)} : \C_n \to \C_n$ and $\Obf^{(2)} : \C_n \to \C_n$.
\item The function $\asset^{(2)}$ is defined on $\Obf^{(1)}(P,\aux)$ and the function $\asset^{(1)}$ is defined on $\Obf^{(2)}(P,\aux)$ for $P \in \C_n$.
(This is essentially implied by (1) and the definition of an asset.)
\item 
Over $P \in \C_n$ generated by $(P, \aux) \leftarrow \Gen_\C(1^n)$, the output distribution of programs $P' \leftarrow \Obf^{(1)}( \Obf^{(2)}(P,\aux),\aux)$ is close\footnote{To be precise, the statistical distance of the distributions is small.} to the output distribution of $\Obf^{(2)}(\Obf^{(1)}(P,\aux),\aux)$.
\end{enumerate}
Then, there is no adversary that can compute either $\asset^{(1)}(P)$ or $\asset^{(2)}(P)$ from $\Obf^{(2)}(\Obf^{(1)}(P,\aux),\aux)$.
\end{theorem}

\begin{proof}
Let $A$ be an adversary that takes as input $\Obf^{(2)}(\Obf^{(1)}(P,\aux),\aux)$ and outputs a candidate $a$ for $\asset^{(1)}(P)$ or $\asset^{(2)}(P)$.
Suppose $A$ succeeds with non-negligible probability (over the probability distribution of $(P,\aux) \leftarrow \Gen_\C( 1^n )$ and the random choices by the obfuscators).

First, consider the case where $A$ computes $\asset^{(1)}(P)$.
Then, we turn $A$ into an adversary $A'$ that computes $\asset^{(1)}(P)$ from $\Obf^{(1)}(P,\aux)$, thus violating the claim that $\Obf^{(1)}$ is a secure obfuscator.
The construction is quite simple: Given a challenge $P' = \Obf^{(1)}(P,\aux)$ the new adversary $A'$ simply applies $\Obf^{(2)}$ to $P'$ and then passes this to $A$, which computes $\asset^{(1)}(P)$ with non-negligible probability.

Next, consider the case where $A$ computes $\asset^{(2)}(P)$.
We turn $A$ into an adversary $A''$ that computes $\asset^{(2)}(P)$ from $P'' = \Obf^{(2)}(P,\aux)$.
The main idea is to use property (3). We apply $\Obf^{(1)}$ to $P''$, to get a program $P'''$.
Now if the distribution of $P''' = \Obf^{(1)} ( \Obf^{(2)}(P,\aux),\aux )$ has small statistical distance from the distribution of $\Obf^{(2)}(\Obf^{(1)}(P,\aux),\aux)$ then the algorithm $A$ also succeeds on $P'''$ with non-negligible probability.
Once again, $A''$ simply returns the value computed by $A$ on this obfuscated program. 
Since $A$ computes $\asset^{(2)}(P)$ with non-negligible probability, it follows that $A''$ also succeeds with non-negligible probability. \qed
\end{proof}

\section{Limitations of Our Formalism}\label{sec:limitations}

We have discussed a number of possible application scenarios for obfuscation, and explained how to use our formalism to define security in each of these settings.
We now discuss some possible objections or questions regarding our formalism.

\subsection{Choosing the Right Asset}

A developer might think they have defined their asset adequately, but have failed to anticipate an attack that computes some related information or partial information.
There are many examples in cryptography where a small amount of ``innocent looking'' leakage can be leveraged to a full attack.
Alternatively, it may be the case that some sensitive data is used in more than one program segment, in which case protecting only one of the segments with strong obfuscation is not going to prevent an attacker from learning the data by attacking one of the other segments.

We agree with these objections.
Our formalism does require the developer to be able to understand where the asset is being used in a program and to anticipate attacks.
In contrast, stronger notions such as VBB obfuscation do not have this weakness: they ensure that no partial information of any type is leaked.

We believe this is a necessary compromise to obtain practical systems. 
In practice, we believe that the task of defining an asset in a program will clarify the security risks and make it clear to the developer in what ways the asset is visible to an attacker within the software.

\subsection{Is an Asset Verifier Necessary?}


The security definitions require the \emph{existence} of an efficient asset verifier algorithm, in order to be able to define the success of an attacker. 
However, this is really a thought experiment. 
It suffices that it would be possible to implement an efficient asset checker. 
It is not necessary for the obfuscated program itself to contain code for this verifier.
On the other hand, when setting up an automated obfuscation challenge it is necessary to actually implement a verifier.

However, it is worth stressing that the verifier is an important feature of our formalism. It allows is the success of an attacker to be explicitly defined. 
A weakness of some previous work on practical obfuscation is that it is unclear what it would mean to ``break'' the obfuscation.

\subsection{The Challenge of Control Flow Obfuscation}

Many papers \cite{Wang-DSN01-Soft-Survival,Majumdar-ICISS06-CFG-Survey,Balachandran-TIFS13-CFG} in the software security community are about tools to make it hard to determine the control flow graph of the original unobfuscated program. 
Understanding/simplifying control flow is an important first step towards ``comprehension'' of a program. 

\draft{
Note that an attacker can try to determine control flow using symbolic execution, or actual execution traces, or by considering data flow and dependencies, or many other methods. 
So, successfully hiding control flow seems to also require hiding many aspects of a program that, at first sight, may not seem to be about control flow. 
As such, it is unclear whether specifying control flow as the asset to be protected is actually narrowing-down the scope of obfuscation very much. 
There may be little difference between a special-purpose obfuscator to hide control flow and a general-purpose obfuscator that hides all aspects of a program.
}

We now explain some of the complexities of trying to express control flow obfuscation in our formalism:

\begin{enumerate}
\item It is necessary to specify a class $\C$ of programs that have ``rich'' control flow. 
It is not obvious how to make this precise or how to give a compact representation for such a class.

\item A natural approach is to define $\asset(P)$ to be the Control Flow Graph (CFG) of the code produced by the software developer. 
The obfuscator should output a program $\OO(P)$ with a more complex CFG.

But our formalism requires an efficient algorithm $\verify$  that determines if an attacker has been able to learn the asset.
For CFGs, it is natural to propose that the adversary wins if it can compute a CFG from $\OO( P)$ that is ``topologically close'' to the CFG of the unobfuscated program.
It follows that the program generator $\Gen$ should output as auxiliary data the CFG of the original program $P$.
In this case, the verify algorithm is required to determine whether the output CFG is ``close'' to the original CFG.
It is challenging to make this precise and to understand the complexity of the verification algorithm.
Chan and Collberg~\cite{CC14} have proposed a method based on edit distance, which is known to be a hard problem \cite{Gao-PAA10-Edit-Distance-Survey}.
\end{enumerate}

Rather than showing a weakness of our formalism, we believe that our formalism gives insight into the difficulty of control flow obfuscation. We have explained that even giving a satisfactory definition for security of control flow obfuscation is hard. 
Through this lens, it becomes less surprising that attempts to create general-purpose obfuscation tools for control flow obfuscation have not been successful.


\subsection{Types of Asset}

At a high-level, there are two broad types of program that are often mentioned in high-level discussions of program obfuscation.
\begin{enumerate}

\item The first type of programs contain some specific data (\eg a licence key, a password, a cryptographic key, or a solution to some computational problem that is hard to find), which is used/accessed in the program but that we wish to keep secret.
The programs themselves might be quite simple, and their functionality might be fully known to an attacker.
In this class of programs, the asset itself is clearly defined (though it must come from a large set in order not to be guessable). 
White-box cryptography (see Section~\ref{sec:WBC}) is one such type of obfuscation, where the class of programs is a symmetric cipher with an arbitrary key.
The majority of theoretical work on obfuscation (\eg \cite{BBCKPS14}) is about this first type.

\item For the second type, there is some kind of intellectual property of the software developer that is encapsulated in the program code.
If a program $P$ is easily learnable from viewing its input-output behaviour, or if the functionality of $P$ is clearly specified such that it is easy for human software developers to write their own program $P'$ to do the same task, then there is no point in obfuscating the program.
Hence, when discussing the second type of programs, the class has to be rich enough to include a wide variety of real-world programs that would require significant software development time.
Two examples of such program classes are evasive finite automata~\cite{GaZo20} and compute-and-compare programs~\cite{WZ17}, though these both have something of the character of the first type of program class as well.
In the formalism, we require that the asset is not learnable from black-box access to the program.
The precise definition of an asset can be harder to define for these types of programs in general, and we believe that the difficulty of formalising the problem is one of the impediments to progress in practical obfuscation. We give some examples in Sections~\ref{sec5.5}, \ref{sec5.6}, \ref{sec5.8}, \ref{sec5.9}.
\end{enumerate}

It is notable that in the entire literature, including this paper, the focus tends to be on the first situation. 
One can speculate that the difficulty of defining ``intellectual property of software'' is yet another obstacle to be overcome in order to develop a complete theory of program obfuscation.

\section{Conclusion}
\label{sec:conclusion}

We have introduced a new formalism that allows to precisely define security of obfuscation and to rigorously prove the security of an obfuscator.
We believe our formalism will be useful for the design (and cryptanalysis) of obfuscation tools.

Central to our formalism is the notion of an asset.
Our contributions include clarifying the need for a precise definition of program class (via a program generator) and the requirement to have an asset verifier.
We have shown that these requirements are exactly the same requirements for setting up an automated obfuscation challenge.

\section*{Acknowledgment}

We thank Christian Collberg for several suggestions.
This research is funded in part by the Royal Society of New Zealand Marsden fund project 16-UOA-144.


\bibliographystyle{IEEEtran} 
\bibliography{obfuscation}

%
%

%

\end{document}